

\magnification=1200
\hsize=6.4 truein \vsize=8.9 truein

\tolerance=10000
\def\folio{\ifnum\pageno=1\nopagenumbers\else\number\pageno\fi}
%
        \def\footstrut{\baselineskip=10pt plus 1pt}
        \skip\footins=30pt
        \def\footnoterule{\kern-3pt
        \hrule width \the\hsize \kern 2.6pt}
\baselineskip 12pt minus 1pt
\parskip=\medskipamount
\def\nulltest{}    
\def\headline{}    
\def\headlineb{}   
\def\makeheadline{\ifnum\pageno=1 \else
  \baselineskip 12pt \ifx \headline \nulltest \else
     \line{\headline} \advance \vsize by -\baselineskip \advance
     \vsize by -14pt \fi
   \ifx \headlineb \nulltest \else
     \line{\headlineb} \advance \vsize by -\baselineskip \fi
   \ifx \headline \nulltest \else \vskip 14pt \fi
  \fi}   
\def\page{\vfill\eject}
\def\undertext#1{$\underline{\smash{\hbox{#1}}}$}
\def\etal{{\it et al.\ }}
\def\refpar{\parshape=2 0truein 6.5truein 0.3truein 6.2truein}
\def\ref#1;#2;#3;#4{{\par\refpar #1, {\it #2}, {\bf #3}, #4}}
\def\prep#1;#2{{\par\refpar #1, #2}}
\def\book#1;#2;#3{{\par\refpar #1, {\it #2} #3}}
\def\SH#1{\vskip 12pt\goodbreak\centerline{\bf#1}}
\def\ltsima{$\; \buildrel < \over \sim \;$}
\def\lsim{\lower.5ex\hbox{\ltsima}}
\def\gtsima{$\; \buildrel > \over \sim \;$}
\def\gsim{\lower.5ex\hbox{\gtsima}}
\def\ergcms{\,{\rm erg\,cm^{-3}\,s^{-1}}}
\def\kms{\,{\rm km\,s^{-1}}}
\def\gcc{\,{\rm g\,cm^{-3}}}
\def\pcc{\,{\rm cm^{-3}}}
\def\kel{\,{\rm K}}
\def\grad{\nabla}
\def\NVC{$~N(V_{\rm circ})\Delta V_{\rm circ}~$}
\def\VC{$~V_{\rm circ}~$}
\def\ltsima{$\; \buildrel < \over \sim \;$}
\def\lsim{\lower.5ex\hbox{\ltsima}}
\def\gtsima{$\; \buildrel > \over \sim \;$}
\def\gsim{\lower.5ex\hbox{\gtsima}}
\def\Mpc{\rm Mpc}
\vbox to 0.25in{}
\line{\hfill FERMILAB-Pub-92/217-A}
\line{\hfill UCLA/92/TEP/27}
\line{\hfill August, 1992}
\vbox to 0.5in{ }
\centerline{\bf Large- and Small-Scale Constraints on Power Spectra}
\centerline{\bf in $\Omega=1$ Universes}
\vskip .15truein
\centerline{\bf James M. Gelb$^1$, Ben-Ami Gradwohl$^{1,2}$,
and Joshua A. Frieman$^1$}
\vskip .15truein
\centerline{$^1$NASA/Fermilab Astrophysics Center}
\centerline{Fermi National Accelerator Laboratory}
\centerline{P.O. Box 500, Batavia, IL 60510}
\vskip .15truein
\centerline{$^2$Department of Physics}
\centerline{University of California Los Angeles}
\centerline{Los Angeles, CA 90024}
\vskip .25truein
\centerline{\bf ABSTRACT}
\vskip .1truein
{\parindent 0pt
The cold dark matter (CDM) model
of structure formation, normalized on large scales, leads to excessive
pairwise velocity dispersions on small scales.
In an attempt to
circumvent this problem,
we study three scenarios (all with
$\Omega=1$)
which have more large-scale power and less
small-scale power than the CDM model:
1) an admixture of cold and hot dark matter;
2) cold dark matter with a non-scale-invariant,
power-law primordial power spectrum; and
3) cold dark matter
with coupling of dark matter to a long-range vector field.
Despite the {reduced} small-scale
power, when
such models are evolved in the nonlinear regime
to large amplitude,
the velocities on small scales are actually
{\it increased} over CDM with the same value of $\sigma_8$.
This `flip-over', in disagreement with
the expectation from linear perturbation
theory, arises from
the nonlinear coupling of the extra power on large scales
with shorter wavelengths.  However, due to the extra
large-scale power, the recent COBE DMR
results indicate smaller amplitudes
for these models, $\sigma_8 \sim 0.5 - 0.7$, than for CDM
(for which $\sigma_8 \sim 1.2$).  Therefore, when normalized to
COBE on large scales, such models do lead to reduced velocities
on small scales and they produce fewer
halos compared with CDM. Quantitatively it seems, however,
that models that produce sufficiently low small-scale
velocities fail to produce an adequate distribution of halos.
}
\vskip .25truein
\centerline{\it Submitted to Astrophysical Journal Letters}
\vfill
\eject
\baselineskip 0.6cm {
\vskip .15truein
\centerline{\bf 1. INTRODUCTION}

Recent observations
of fluctuations in
the cosmic microwave background by the COBE satellite (Smoot et al. 1992) have
placed the
gravitational instability of dark matter theory of structure formation
on firmer footing.  At the same time, the shape of the linear
density fluctuation
spectrum
is now constrained by a variety of observations ranging from galaxy
scales ($\sim$ Mpc) up to the very large scales ($\sim $ 1000 Mpc)
probed by
COBE.  The most popular theory of galaxy formation, the
cold dark matter model with a scale-invariant spectrum of
primordial fluctuations
(hereafter CDM),
has been studied by many
authors using numerical $N$-body experiments over a wide range of scales
(e.g. Davis et al. 1985, hereafter DEFW; Melott 1990; Park 1990;
Couchman \& Carlberg 1992; Gelb 1992, hereafter G92).
Using the COBE DMR measurements to fix the density fluctuation
amplitude on large scales, it is now necessary to re-evaluate the
predictions of the CDM model on smaller scales.  In fact,
it is difficult to reconcile the COBE measurements, which
indicate a rather high amplitude for CDM,
with the observed low galaxy pairwise velocity dispersions on scales
of a few Mpc.  In a search for solutions to this problem,
we consider three scenarios, each of which might be considered
a variation on the CDM model and has less small-scale and more
large-scale power than CDM.

Given current theoretical uncertainties,
the amplitudes of the initial power spectra for the models
are free parameters; they are
related, for example, to the self-coupling of the scalar field
that drives inflation.  It is common to normalize spectra by
setting the {\it rms} density fluctuation
in spheres of radii 16 Mpc to $\sigma_8$, assuming
linear growth of modes independent of
their wavenumber $k=2\pi/\lambda$.
(We assume $H_0=50\kms~{\rm sec}^{-1}$
and $\Omega=1$ throughout.)  Thus, for a power spectrum $P(k)$,
\vbox{
$$\sigma_8^2\equiv {\intop_0^\infty} d^3k P(k)W_{\rm TH}^2(kR)~;~~
W_{\rm TH}(kR) = {3\over{(kR)^3}}(\sin kR - kR \cos kR)~;~~R=16~{\rm Mpc}~.
\eqno(1)$$
}
On small scales, $\sigma_8$ is constrained by the pairwise velocity dispersion,
$\sigma_{\parallel}$, defined as the
{\it rms} velocity along lines of separation of galaxy pairs:
$\sigma_{\parallel}(r)\equiv{\langle ( v_{||}-{\langle v_{||}\rangle} )^2
\rangle}^{1/2}$, where $v_\parallel= (\vec v_2-\vec v_1)\cdot \vec r/r$
is the parallel, peculiar velocity and
$\vec r\equiv \vec r_2-\vec r_1$.

The pairwise velocity dispersion of galaxies on small scales
was determined from redshift surveys by Davis \& Peebles (1983):
$\sigma_\parallel ~(r \lsim 3~ \Mpc) \simeq 300\pm 50\kms$
(Bean et al. 1983 found $250\pm 50\kms$).
Researchers have compared this value
with $N$-body simulations
in order to constrain the CDM model amplitude $\sigma_8$.
DEFW found $\sigma_8=0.4$ and G92 found
$\sigma_8\lsim 0.5$ (allowing for galaxy dispersions in simulated clusters
to be less than virial estimates).
However, the recent COBE measurements imply a larger
amplitude for the CDM model: the observed
{\it rms} fluctuation on $10^\circ$,
$[\sigma_T(10^\circ)]_{\rm DMR} = (1.085\pm 0.183)\times 10^{-5}$,
sets $\sigma_8 \simeq 1.17 \pm 0.23$ [Adams et al. 1992; a nearly identical
range for $\sigma_8$ is obtained by fitting the COBE angular correlation
function $C(\theta)$].  A larger amplitude
($\sigma_8 \simeq 1.29[1^{+.38}_{-.65}]$, Adams et al. 1992)
is also required for CDM by
the large-scale peculiar velocities of galaxies,
determined by Bertschinger et al. (1990) using the
POTENT reconstruction algorithm.
Thus, for CDM, large-scale observations suggest a $\sigma_8$ amplitude
roughly {\it twice as large} as that indicated by $\sigma_\parallel$ on small
scales.

A possible way to reconcile large scales with small scales was
suggested by
Couchman \& Carlberg (1992), who
found that $\sigma_\parallel$ for the resolved halos is a factor
$\sim 2$ lower than $\sigma_\parallel$ for the mass, an effect
known as `velocity bias'.
However, G92 showed that
despite a velocity bias of
$b_v\equiv \sigma_\parallel({\rm halos})/\sigma_\parallel({\rm mass})
\sim 1/2$,
high amplitude CDM still produces $\sigma_\parallel$ on scales of
$r > 1$ Mpc in excess of the observed $\sigma_\parallel$.
Furthermore, CDM simulations
result in an overproduction of halos, with the high-mass halos
(from mergers) possibly representing clusters. If so, then dividing up the
clusters into halos eliminates the velocity bias altogether
(see G92). The alternative approach of using peak
particles to represent galaxies (Bardeen et al. 1986) also leads to
$b_v\simeq 1$ (DEFW; Quinn, Katz, \& Gelb 1992).
We therefore claim, that $\sigma_\parallel({\rm mass})$ adequately reflects
$\sigma_\parallel({\rm halos})$, and consequently only focus on
$\sigma_\parallel({\rm mass})$ in the following analysis.
\vskip .15truein
\centerline{\bf 2. MODELS}

We present results from $128^3$ particle
simulations using the
particle-mesh technique on a $256^3$ grid
(Bertschinger \& Gelb 1991) in boxes of comoving length
200 Mpc on a side.  This size
is large enough
to encompass the waves that
contribute significantly to
$\sigma_8$ and $\sigma_\parallel$, although
these simulations, with particle mass
$m_p=2.6\times 10^{11}{\rm M}_\odot$,
are inadequate for resolving individual halos.
(We use 51.2 Mpc boxes, $m_p=4.4\times 10^9{\rm M}_\odot$,
at the end of $\S4$ to study halo formation.)

We explore three models with less small-scale power
and more large-scale power than the CDM model:
(1) models with an admixture of hot and cold dark matter;
(2) a cold dark matter model
with a non-scale-invariant, power-law primordial
spectrum, $P(k) \propto k^n$, with $n = 0.7$;
and (3) cold dark matter models in which the
dark matter couples to a hypothetical
long-range field (Frieman \& Gradwohl 1991;
Gradwohl \& Frieman 1992).
In all cases, we assume negligible
baryon density, and, with the exception of (2),
a scale-invariant ($n=1$) primordial spectrum.  In case (3),
the force law between objects of mass $m$ is:
$$r^2F/(Gm^2)=1 + \alpha (1+r/\lambda_0)\exp(-r/\lambda_0)~,\eqno(2)$$
so that gravity is
effectively retarded for $\alpha < 0$ in the case of a vector field.
Here, $\alpha$ is a measure of the relative strength of the new force,
and $\lambda_0$ is its range.  For these `alpha' models,
we regenerate
the optimal Green's function (Hockney \& Eastwood 1982)
for the potential corresponding
to the force law of eqn.~2,
$$\hat\phi\propto{\left[{ {1\over {k^2}}+{\alpha\over{k^2+(2\pi/\lambda)^2}}
}\right]}{\hat\delta}~;\eqno(3)$$
$\lambda=\lambda_0 a_0/a$ is the comoving scale
at expansion factor $a$, and $\hat\delta$ is the
Fourier transform of the density contrast.
(We define $a=1$ when $\sigma_8=1$ throughout.  The comoving scale
equals the physical scale at present day expansion factor $a_0=\sigma_8$.)
Because of limited force resolution,
the retardation effect is washed out
at later expansion times --- the principle effect enters
in the initial conditions (Gradwohl \& Frieman 1992) and the early evolution.

Other authors have studied some of these models.
Cen et al. (1992) simulated a $\sigma_8=0.5$ $n=0.7$ cold dark matter model
with comparable resolution.
The authors argued that this model has lower
$\sigma_\parallel
\sim 400-500 \kms$
than $\sigma_8=1$ CDM, but that a velocity bias $b_v\simeq 1/1.5$
is still needed
to match the observed $\sigma_\parallel\sim 300\kms$.
However, we do not allow a velocity bias factor
for reasons discussed earlier.
Davis, Summers, \& Schlegel (1992) performed high resolution
simulations with cold and hot dark matter ($\Omega_{\rm HDM}=0.30$)
in 14 Mpc boxes at $\sigma_8=0.9$.  We demonstrate
that not only is this box too small to adequately measure $\sigma_\parallel$
(as pointed out by the authors),
but that large-scale waves
(not present in a 14 Mpc box) at large values of $\sigma_8$
can actually increase
$\sigma_\parallel$ on small scales
relative to CDM.
\vskip .15truein
\centerline{\bf 3. LINEAR PERTURBATIONS}

We present linear realizations of the initial power spectra
and linear estimates of $\sigma_\parallel$
for the models.
The dimensionless
linear power
spectra $P(k)(\Delta k)^3$ (all with $\sigma_8=1$) are plotted
in fig.~1a
as a function of comoving wavenumber $k$,
where $\Delta k=(2\pi/L)^3$ with $L=200$ Mpc.
We begin all simulations at an expansion
factor $a=1/50$.
We use the transfer function of DEFW for models
other than the cold plus hot models,
and the density-weighted transfer function
of van Dalen \& Schaefer (1991) for the cold plus hot models.
All of the simulations use the same initial random numbers
scaled to the appropriate power spectra.
Results are shown for six simulations:
CDM (standard cold dark matter);
C+H13 and C+H30 (cold dark matter mixed with 13\%  and 30\% hot dark matter);
TILT7 (cold dark matter with $n=0.7$); and
ALPHA3 and ALPHA5 (cold dark matter
alpha models with $\alpha=-0.3$, $\lambda_0=100$ kpc;
and $\alpha=-0.5$, $\lambda_0=500$ kpc).
The sharp cutoff corresponds to
the three-dimensional Nyquist wavenumber and
the solid line is a direct plot of the DEFW
CDM power spectrum for discrete $kL/(2\pi)$.
The first bin for the simulated models is an average over 18 waves.

The TILT7, C+H13, and ALPHA3 power spectra
are similar.  However, the alpha simulations will lag
behind the others as the modified force law continues
to retard the subsequent evolution.
(We discuss the nonlinear power spectra of fig.~1b
in $\S 4$.)

Although it is a poor approximation for the small scales of interest,
it is nevertheless instructive to estimate
$\sigma_\parallel$ in linear perturbation theory where
the velocity and density are related by
(Peebles 1980; Lightman
\& Schechter 1990)
$$\vec\nabla\cdot\vec v\approx -H_0\Omega^{4/7}\delta~,\eqno(4)$$
where $\delta$ is the density contrast;
therefore, $v^2_kd^3k \propto H_0^2P(k)/k^2d^3k$.  As a result, the
linear estimate is
$$\sigma_\parallel^2(r)\equiv \langle|{\vec v}(r)-{\vec v}(0)|^2\rangle=
2[\langle v^2\rangle-\langle {\vec v}(r)\cdot{\vec
v}(0)\rangle]$$
$$~~~~~~~~~=2H_0^2\intop_{2\pi/\lambda_{\rm max}}^\infty {4
\pi k^2 dk}{P(k)\over k^2}\left [1-{{\sin kr}\over{kr}}
\right]~.\eqno(5)$$
\vfill
\eject
{\parindent 0pt The factor
$1-{{\sin (kr)}/{(kr)}}$ filters out the contribution of long
waves to the small-scale pairwise velocity dispersion;
for waves with $\lambda\gg r$, the bulk flow does not contribute
to the dispersion.  This is opposite to the
top hat filter $W_{\rm TH}$ in eqn.~1, which filters out
contributions from large $kr$.}

In fig.~2a we plot linear theory estimates of
$\sigma_\parallel$ for the various spectra
at $\sigma_8=1$.  The Davis \& Peebles (1983) estimates
from the observations are $\sim 300\pm 50\kms$
for scales $\sim 1-3$ Mpc.  The models with reduced small-scale
power appear to fare well at these scales, but, as we now show,
nonlinear effects radically alter $\sigma_\parallel$.
\vskip .15truein
\centerline{\bf 4. NONLINEAR CALCULATIONS}

\vskip .15truein
\centerline{\bf 4.1 PAIRWISE VELOCITY DISPERSIONS}

We first study nonlinear power spectra and $\sigma_\parallel$
for the models.
In fig.~1b we present
nonlinear power spectra at $\sigma_8=1$ for several models,
computed from our simulations in the 200~Mpc box.
For low values of $k$, the spectra
agree with their counterparts in
fig.~1a, indicating a sufficiently large box size.
The models with less small-scale power in the initial
conditions continue to have less small-scale power in the nonlinear regime,
yet the nonlinear spectral shapes differ from the linear
regime
and the differences among the various models are significantly
reduced, compared to our linear estimates.

The nonlinear $\sigma_\parallel$ for the models are shown in figs.~2b and 2c.
Comparing with fig.~2a, it is clear that linear theory is a very
poor estimator of both the amplitude and the general characteristics
of $\sigma_\parallel$.
In fig.~2b $\sigma_\parallel$ is plotted at $\sigma_8=$0.25, 0.5,
0.75, and 1 for CDM and C+H30.  At
low amplitude, $\sigma_8 \lsim 0.5$, $\sigma_\parallel$ for
C+H30 is
lower than $\sigma_\parallel$ for CDM, in agreement with the expectation from
linear theory. However, when the models are evolved further, i.e.,
for $\sigma_8 \gsim 0.7$, $\sigma_\parallel$ `flips over': despite its
reduced small-scale power, the C+H30 model yields {\it larger}
$\sigma_\parallel$ than CDM.  Fig.~2c shows that this flip-over
at high $\sigma_8$ is generic for these models; it is
a reflection of the fact
that their extra power on large scales
couples significantly to small scales.  This effect is manifest in
the pairwise velocity dispersion, but not the power spectrum,
because the former is
more sensitive to long wavelengths.

The flip-over in the initial power spectra
occurs on scales exceeding $\sim 75$ Mpc (fig.~1a)
and is therefore missed in
the 14 Mpc simulations of Davis, Summers, \& Schlegel (1992); at
$\sigma_8=0.9$ (the case studied there) C+H30
actually yields higher $\sigma_\parallel$
than the CDM model.
(Originally we used 51.2 Mpc boxes
and did not see this effect either.)
Cen et al. (1992) may not have elucidated this effect
since they only evolved their $n=0.7$ simulation to $\sigma_8=0.5$.
Our TILT7 results agree with Cen et al. (1992)
at this amplitude.

To compare these results with the
observations of $\sigma_\parallel
\simeq 300\pm 50\kms$, we fix the $\sigma_8$ amplitudes of
the models using the DMR data and linear perturbation theory
(which is valid on the large scales probed by COBE). As noted in
$\S 1$, for CDM, COBE yields $\sigma_8 \gsim 1$, which
implies $\sigma_\parallel$ in excess of 1200 km s$^{-1}$ over scales of
a few Mpc. For the other models, COBE's {\it rms} fluctuations
on $10^\circ$, including the DMR errors and cosmic variance for
the models, imply $\sigma_8 \simeq 0.78 \pm 0.16$ (for C+H13),
$0.69 \pm 0.14$ (for C+H30), $0.53 \pm 0.11$ (for TILT7),
$0.95 \pm 0.19$ (for ALPHA3),
and $0.51 \pm 0.10$ (for ALPHA5).  When normalized to COBE, it is possible
to find models with reduced $\sigma_\parallel$. One should, however, point
out that models with $\sigma_\parallel\sim 400-550\kms$, although favorable
over
$\sigma_8=1$ CDM, may still be inadequate. This is due to the fact that
simulation-to-simulation variations
in $\sigma_\parallel$ are typically $\lsim 100\kms$ (G92),
the observed errors in $\sigma_\parallel$ are $\sim 50\kms$, and
Bean et al. (1992) found lower $\sigma_\parallel$ of order
$250\pm 50\kms$ than Davis and Peebles (1993).

In fig.~3a we plot $\sigma_\parallel$ for `favorable' models, subject to the
above COBE normalizations. They are:
$\sigma_8=0.6$ ALPHA5 and $\sigma_8=0.5$ TILT7,
both consistent with COBE; $\sigma_8=0.5$ C+H30, which is somewhat
beyond the $1\sigma$ level; and
$\sigma_8=0.5$ CDM, which is inconsistent with COBE and only shown here
for comparison.
We exclude C+H13 and ALPHA3 from our list, as they
produce excessive $\sigma_\parallel$.
TILT7 and C+H30 both produce small-scale $\sigma_\parallel$ in the
$400-550\kms$
range, a factor of $\sim 2$ smaller than $\sigma_8=1$ CDM, but still at least
30\% above the observed $\sigma_\parallel$.
The only model in fig.~3a
that matches COBE and small-scale $\sigma_\parallel$ is
ALPHA5. It nicely reproduces the observed $\sigma_\parallel$, but as we
now demonstrate, suffers from inadequate halo formation.
\vskip .15truein
\centerline{\bf 4.2 HALO FORMATION}

In addition to the affect on $\sigma_\parallel$, reducing small-scale power
can also help alleviate some problems
associated with excessive
low-mass halo formation in the field and an excessive number density of
high-mass objects
that plague the CDM model (White et al. 1987;
G92). In order to study halo formation, we simulate models
in 51.2~Mpc boxes, again using
$128^3$ particles and a $256^3$ particle-mesh grid. (We emphasize
again, that although adequate for analyzing the halo distribution, the
51.2~Mpc box simulations {\it cannot} be used to compute $\sigma_\parallel$.)

The details of galaxy formation require
higher resolution simulations
with separate hot and cold dark matter particles
(e.g. Davis, Summers, \& Schlegel 1992)
and separate dark matter and baryonic matter particles
in alpha models (the coupling to the vector field
only occurs for the dark matter and can lead to a natural
bias between dark and baryonic matter, Gradwohl \& Frieman 1992).
These considerations, however, are not likely to significantly affect
mass pairwise velocity dispersions.

Halos in our simulations are identified, by use of
the DENMAX algorithm (Bertschinger \& Gelb 1991),
as local density maxima in the evolved, nonlinear
density field.
The distribution of halos, characterized by their circular velocities
(computed from the enclosed mass within $R=$300 kpc of the
DENMAX center, i.e. $\sqrt{GM(<R)/R}$)
are then counted in $25\kms$ bins.  The results
are shown in fig.~3b for the same scenarios as
in fig.~3a.  The solid line
is the observed estimate, using a Schechter (1976) luminosity
function coupled with
Tully-Fisher (1977) and Faber-Jackson (1976) relationships
(assuming that 70\% of the halos are spirals and 30\% are ellipticals)
to relate luminosity to mass.
(See G92 and White et al. 1987 for
details.  The observed estimates as shown tend to overestimate
the highest-mass halos compared with complete elliptical surveys,
and the overall normalization has an error $\sim 30\%$.)

The cases $\sigma_8=0.5$ CDM and $\sigma_8=0.5$ TILT7 appear to match
the observed distribution fairly well.
G92 demonstrated, however, that high-mass halos
should be divided into clusters
of halos so that 1) the simulations contain clusters and 2)
extra weight is given to dense systems thereby enhancing
the two-point correlation function in
biased ($\sigma_8<1$) models (White et al. 1987).
Therefore, TILT7 actually does much better
than CDM---it produces less mid-mass halos and less high-mass halos
which, when divided into mid-mass cluster members, make up the
mid-mass deficit.
C+H30 produces too few halos,
which can be remedied by evolving the simulation further at
the expense of raising $\sigma_\parallel$.
ALPHA5 drastically fails to match the observed distribution for $\sigma_8=0.6$.
(ALPHA3 does better, but it produces excessive $\sigma_\parallel$.)
\vskip .15truein
\centerline{\bf 5. CONCLUSIONS}

For $\sigma_8 \gsim 0.5$ (the precise value depends on the model),
the nonlinear coupling of waves in models with more large-scale
power and less small-scale power than
CDM actually
{\it increases} $\sigma_\parallel$ on small scales,
in complete disagreement with linear perturbation theory.
At $\sigma_8=0.5$, CDM, C+H30, and TILT7
yield $\sigma_\parallel\sim 400-550\kms$ on small scales, but
CDM  has inadequate large-scale amplitude.
C+H30 and TILT7 also generate, consistent with observations, less
halos than CDM.
The only model which produces $\sigma_\parallel\sim 300\kms$, and at the same
time matches COBE, is ALPHA5 at $\sigma_8=0.6$,
but it fails to produce a sufficient number density of halos.

The fact that it seems difficult (if not impossible?) to accommodate a
low $\sigma_\parallel$, and still have enough small-scale power for adequate
halo formation, may hint to a basic problem of $\Omega=1$ cosmogonies. One way
out of this apparent impasse is, of course, to lower the matter density of the
universe (recall that $v \propto \Omega^{4/7}$), and thereby maintain a low
$\sigma_\parallel$ with increased small scale power. It is clearly too early to
view this problem as a death stroke to $\Omega=1$ scenarios, more work still
needs to be done involving the details of halo formation and biasing.

We summarize our results in Table 1.
There are no obvious `winners'.  The balance of scores can
be shifted by varying $\sigma_8$; e.g.,
C+H30 can be shifted to $+$,$-$,$+$ by choosing
a higher amplitude.
Recent COBE measurements on large scales require the investigation of
a myriad of models with more free parameters
than CDM.  On small scales, in $\Omega=1$ scenarios, it seems
difficult to reconcile
low $\sigma_\parallel$ with sufficient halo formation, and one may be forced to
consider scenarios with $\Omega < 1$ or a cosmological constant.

We thank Ed Bertschinger, Dick Bond,
Bob Schaefer, and Mike Turner for stimulating discussions.
This research was conducted using the Cornell National Supercomputer
Facility, a resource of the Center for Theory and Simulation in
Science and Engineering at Cornell University, which receives major
funding from the National Science Foundation and IBM
Corporation, with additional support from New York State
and members of its Corporate Research Institute.
This work was supported in part by the DOE and
NASA grant NAGW-2381 at Fermilab.  B.G. was also supported by
DOE \#DE-FG03-91ER (40662 Task C) at UCLA.}
\vfill
\eject
{\parindent 0pt
\baselineskip 12pt minus 1pt{
\vskip .15truein
\centerline{\bf REFERENCES:}

\line{Adams, F. C., Bond, J. R., Freese, K., Frieman, J. A.,
\& Olinto A. V. 1992, preprint\hfill}
\line{Bardeen, J. M., Bond, J. R., Kaiser, N., \& Szalay, A. S. 1986,
{ApJ}, {300}, 15\hfill}
\line{Bean, A. J., Efstathiou, G., Ellis, R. S., Peterson, B. A.,
\& Shanks, T. 1983,\hfill}
\line{~~~~~MNRAS, 205, 605\hfill}
\line{Bertschinger, E., Dekel, A., Faber, S. M., Dressler, A.,
\& Burstein, D. 1990, ApJ, 364, 370\hfill}
\line{Bertschinger, E. \& Gelb, J. M. 1991, Computers in Physics, 5, 164\hfill}
\line{Cen, R., Gnedin, N. Y., Kofman, L. A., \& Ostriker, J. P. 1992,
preprint\hfill}
\line{Couchman, H. M. P. \& Carlberg, R.  1992, ApJ, 389, 453\hfill}
\line{Davis, M., Efstathiou, G., Frenk, C. S., \& White, S. D. M. 1985,
{ApJ}, {292}, 371 (DEFW)\hfill}
\line{Davis, M. \& Peebles, P. J. E. 1983, {ApJ}, {267}, 465\hfill}
\line{Davis, M., Summers, F. J., \& Schlegel, D. 1992, preprint\hfill}
\line{Faber, S. M. \& Jackson, R. E. 1976, {ApJ}, {204}, 668\hfill}
\line{Frieman, J. A. \& Gradwohl, B. 1991, Phys. Rev. Lett.,
67, 2926\hfill}
\line{Gelb, J. M. 1992, M.I.T. Ph.D. thesis (G92);\hfill}
\line{~~~~~also Gelb, J. M. \& Bertschinger, E. 1992, in preparation;\hfill}
\line{~~~~~and Gelb, J. M. 1992, to appear in proc. of
Groups of Galaxies Workshop,\hfill}
\line{~~~~~Space Telescope Science Institute, preprint\hfill}
\line{Gradwohl, B. \& Frieman, J. A. 1992, Ap.J., preprint\hfill}
\line{Hockney, R. W. \& Eastwood, J. W. 1982,
Computer Simulation Using Particles\hfill}
\line{~~~~~(New York: McGraw-Hill)\hfill}
\line{Lightman, A. \& Schechter, P. L. 1990, ApJS, 74, 831\hfill}
\line{Melott, A. L. 1990, Physics Reports, 193, 1\hfill}
\line{Park, C. 1990, {MNRAS}, {242}, 59\hfill}
\line{Peebles, P. J. E. 1980, The Large-Scale Structure of the Universe\hfill}
\line{~~~~~(New Jersey: Princeton University Press)\hfill}
\line{Quinn, T., Katz, N., \& Gelb, J. M. 1992, in preparation\hfill}
\line{Schechter, P. L. 1976, {ApJ}, {203}, 297\hfill}
\line{Smoot, G. F. et al. 1992, preprint\hfill}
\line{Tully, R. B. \& Fisher, J. R. 1977, {Astr. Ap.}, {54}, 661\hfill}
\line{van Dalen, A. \& Schaefer, R. K. 1991, preprint\hfill}
\line{White, S. D. M., Davis, M., Efstathiou, G., \& Frenk, C. S.
1987, {Nature}, {330}, 451\hfill}
\vfill
\eject
\def\endtable{\endgroup}
\def\tableheight{\vrule width 0pt height 8.5pt depth 3.5pt}
{\catcode`|=\active \catcode`&=\active
    \gdef\tabledelim{\catcode`|=\active \let|=\vbar
                     \catcode`&=\active \let&=\nobar} }
\def\table{\begingroup
    \def\twidth{\hsize}
    \def\tablewidth##1{\def\twidth{##1}}
    \def\defaultheight{\vrule width 0pt height 8.5pt depth 3.5pt}
    \def\heightdepth##1{\dimen0=##1
        \ifdim\dimen0>5pt
            \divide\dimen0 by 2 \advance\dimen0 by 2.5pt
            \dimen1=\dimen0 \advance\dimen1 by -5pt
            \vrule width 0pt height \the\dimen0  depth \the\dimen1
        \else  \divide\dimen0 by 2
            \vrule width 0pt height \the\dimen0  depth \the\dimen0 \fi}
    \def\spacing##1{\def\defaultheight{\heightdepth{##1}}}
    \def\nextheight##1{\noalign{\gdef\tableheight{\heightdepth{##1}}}}
    \def\end{\cr\noalign{\gdef\tableheight{\defaultheight}}}
    \def\zerowidth##1{\omit\hidewidth ##1 \hidewidth}
    \def\hline{\noalign{\hrule}}
    \def\skip##1{\noalign{\vskip##1}}
    \def\bskip##1{\noalign{\hbox to \twidth{\vrule height##1 depth 0pt \hfil
        \vrule height##1 depth 0pt}}}
    \def\header##1{\noalign{\hbox to \twidth{\hfil ##1 \unskip\hfil}}}
    \def\bheader##1{\noalign{\hbox to \twidth{\vrule\hfil ##1
        \unskip\hfil\vrule}}}
    \def\spanloop{\span\omit \advance\mscount by -1}
    \def\extend##1##2{\omit
        \mscount=##1 \multiply\mscount by 2 \advance\mscount by -1
        \loop\ifnum\mscount>1 \spanloop\repeat \ \hfil ##2 \unskip\hfil}
    \def\vbar{&\vrule&}
    \def\nobar{&&}
    \def\hdash##1{ \noalign{ \relax \gdef\tableheight{\heightdepth{0pt}}
        \toks0={} \count0=1 \count1=0 \putout##1\end
        \toks0=\expandafter{\the\toks0 &\end} \xdef\piggy{\the\toks0} }
        \piggy}
    \let\e=\expandafter
    \def\putspace{\ifnum\count0>1 \advance\count0 by -1
        \toks0=\e\e\e{\the\e\toks0\e&\e\multispan\e{\the\count0}\hfill}
        \fi \count0=0 }
    \def\putrule{\ifnum\count1>0 \advance\count1 by 1
 \toks0=\e\e\e{\the\e\toks0\e&\e\multispan\e{\the\count1}\leaders\hrule\hfill}
        \fi \count1=0 }
    \def\putout##1{\ifx##1\end \putspace \putrule \let\next=\relax
        \else \let\next=\putout
            \ifx##1- \advance\count1 by 2 \putspace
            \else    \advance\count0 by 2 \putrule \fi \fi \next}   }
\def\tablespec#1{
    \def\vdimens{\noexpand\tableheight}
    \def\tabby{\tabskip=0pt plus100pt minus100pt}
    \def\r{&################\tabby&\hfil################\unskip}
    \def\c{&################\tabby&\hfil################\unskip\hfil}
    \def\l{&################\tabby&################\unskip\hfil}
    \edef\templ{\noexpand\vdimens ########\unskip  #1
         \unskip&########\tabskip=0pt&########\cr}
    \tabledelim
    \edef\body##1{ \vbox{
        \tabskip=0pt \offinterlineskip
        \halign to \twidth {\templ ##1}}} }
$$
\table
\tablespec{\c\c\c\c\c}
\body{
\header{TABLE~1}
\skip{5pt}
\header{Scorecard}
\skip{5pt}
\hline
\skip{2pt}
\hline
\skip{2pt}
&Model &$\sigma_8$&COBE&$\sigma_\parallel(r\lsim 3~{\rm Mpc})$&Halos&\end
\skip{2pt}
\hline
\skip{2pt}
&CDM&0.5&$-$&$\circ$&$-$&\end
&C+H30&0.5&$+$&$\circ$&$\circ$&\end
&ALPHA5&0.6&$+$&$+$&$-$&\end
&TILT7&0.5&$+$&$\circ$&$+$&\end
\skip{2pt}
\hline
}
\endtable
$$
{\parindent 0pt{
\bf Table~1 Caption:} Scorecard for the various models with
$+$ indicating a favorable score, $\circ$ indicating a marginal score,
and $-$ indicating a disfavorable score.}
\vfill
\eject
\centerline{\bf FIGURE CAPTIONS:}
{\bf Figure~1:} a) Linear realizations (all use equivalent random
numbers) of various
power spectra normalized to $\sigma_8=1$.
(C+H30 has less small-scale power than C+H13 and
ALPHA5 has less small-scale power than ALPHA3.)
b) Nonlinear power spectra
at $\sigma_8=1$
for some of the simulations shown in
fig.~1a. In both a) and b),
the solid curve is
the analytic form of the DEFW linear CDM power spectrum.

{\bf Figure~2:}
a) Linear theory calculations of $\sigma_\parallel$
versus comoving separation (eqn.~5) for $\sigma_8=1$.
(C+H30 is lower than C+H13 and ALPHA5 is lower than ALPHA3.)
b) Nonlinear calculations
of $\sigma_\parallel$
for CDM (short dashed curves) and for
C+H30 (long dashed curves) at four values of $\sigma_8$.
c) Nonlinear calculations of $\sigma_\parallel$ for
CDM (short dashed curves), C+H13 (long dashed curves), ALPHA3 (dot-short dashed
curves), and TILT7 (dot-long dashed curves)
at $\sigma_\parallel=$ 0.5 and 1.  The observed
estimates from galaxies for $r\sim 1-3$ Mpc are $300\pm
50\kms$ (Davis \& Peebles
1983) and $250\pm 50\kms$ (Bean et al. 1983).

{\bf Figure~3:} a) Nonlinear calculations (200 Mpc box)
of $\sigma_\parallel$.  b) Distribution of halos versus
circular velocity (in $\Delta V_{\rm circ}=25\kms$ bins).  The solid curve is
the observed estimate.
In both a) and b) the cases are:
CDM at $\sigma_8=0.5$ (short dashed curves),
C+H30 at $\sigma_8=0.5$ (long dashed curves), ALPHA5 at $\sigma_8=0.6$
(dot-short dashed curves),
and TILT7 at $\sigma_8=0.5$ (dot-long dashed curves).
}
\vfill
\eject
\end